\documentclass[%
 reprint,
superscriptaddress,
 amsmath,amssymb,
 pra
]{revtex4-2}

\usepackage{braket}
\usepackage{graphicx}
\usepackage{dcolumn}
\usepackage{bm}
\usepackage{hyperref}

\newcommand{\ep}[1]{#1}
\begin{document}

\preprint{APS}

\title{Quantum  $1/f^\eta$ Noise Induced Relaxation in the Spin-Boson Model}

\author{Florian Otterpohl}
    \email{florian.otterpohl@physik.uni-hamburg.de}
\affiliation{%
I. Institut f\"ur Theoretische Physik, Universit\"at Hamburg, Notkestr. 9, 22607 Hamburg, Germany\\
}%
\author{Peter Nalbach}
\affiliation{
 Fachbereich Wirtschaft \& Informationstechnik, Westf\"alische Hochschule, M\"unsterstr. 265, 46397 Bocholt, Germany
}%
\author{Elisabetta Paladino}
\affiliation{
Dipartimento di Fisica e Astronomia "Ettore Majorana", Università di Catania, 
Via Santa Sofia 64, 95123 Catania, Italy
}%
\author{Giuseppe A. Falci}
\email{giuseppe.falci@unict.it}
\affiliation{
Dipartimento di Fisica e Astronomia "Ettore Majorana", Università di Catania, 
Via Santa Sofia 64, 95123 Catania, Italy
}%
\author{Michael Thorwart}%
\email{michael.thorwart@uni-hamburg.de}
\affiliation{%
 I. Institut f\"ur Theoretische Physik, Universit\"at Hamburg, Notkestr. 9, 22607 Hamburg, Germany\\
}%

\date{\today}

\begin{abstract}
We extend the spin-boson model of open quantum systems to the regime of quantum  $1/f^\eta$ noise characterized by negative exponents of its spectral distribution. Using the numerically exact time-evolving matrix product operator, we find the dynamic regime diagram, including pseudocoherent dynamics controlled by quantum  $1/f^\eta$ noise. We determine the dephasing rate and find for it an empirical formula valid at zero temperature. The bath reorganization energy depends on the infrared bath cutoff frequency, revealing an increased sensitivity of the dephasing on the measurement time of an experiment. \ep{Our results apply to a qubit as an elementary building block of a quantum computer and pave the way towards a quantum treatment of low-frequency noise in more complex architectures.}
\end{abstract}

\maketitle

\section{Introduction}
Dephasing and relaxation in a quantum system arise from its interaction with environmental fluctuations, with their spectral decomposition being a decisive feature that determines the quantum dissipative dynamics. Understanding and modeling their influence, and subsequently tailoring their impact, is key to many research fields~\cite{Leggett1987, WeissBook}, ranging from chemical reaction dynamics to atomic, molecular, and quantum optical systems, as well as the development of quantum computing devices. 

The spin-boson model serves as a paradigmatic building block for understanding the effect of quantum dissipation and describes a quantum two-state system interacting with harmonic bath degrees of freedom \cite{Leggett1987, WeissBook}. At low temperatures and weak coupling, damped coherent oscillations are observed, while stronger dissipation or high temperatures lead to incoherent decay towards thermal equilibrium. The central characteristic of environmental fluctuations is their spectral distribution $J(\omega) \propto \alpha \omega^s$. \ep{It captures the combined effect of the frequency-dependent coupling strength and the density of states of bath oscillators at a given frequency $\omega$. Commonly, it is} modeled as a continuous function of frequency, with the spectral exponent $s$ and the coupling strength $\alpha$. The regime for $s>0$ (denoted as sub-Ohmic, Ohmic, and super-Ohmic for $0<s<1$, $s=1$, and $s>1$,  respectively) has been widely studied using an impressive collection of available numerical and analytical tools \cite{Leggett1987, WeissBook}. The key questions revolve around the environmental conditions under which different types of dynamical behavior occur. These types include a transition from coherent dynamics at weak to incoherent dynamics at strong dissipation, and a transition from a completely delocalized asymptotic thermal state at weak to a localized asymptotic state at strong dissipation. \ep{The bath spectral density also includes a finite band width characterized by a UV cut-off frequency $\omega_c$.} The \ep{resulting} finite response time of the environmental modes may even generate a pseudocoherent system dynamics \cite{otterpohl_hidden_2022} that is \ep{entirely} controlled by the bath parameters \ep{$\alpha, s$ and $\omega_c$}. Such a remaining pseudo-coherence of the system is surprising for sub-Ohmic baths \cite{Chin06,Anders07,Winter09,Alvermann09,Nalbach10, Wang10, Chin11, Kast13, Nalbach13, Schroeder16,Duan17,otterpohl_hidden_2022,Gelin23}, where \ep{low-frequency bath modes couple stronger to the two-level system than in the case of a pure Ohmic bath. While relaxation between the two states is dominantly influenced by the spectral weight of the bath around the frequency of the transition, dephasing between the two states is determined by the low-frequency distribution of the bath modes.}

A wide class of nanosystems \cite{Falci2024} \ep{and, in particular, current solid-state implementations of quantum computing architectures} \cite{Paladino2002, AstafievPRL2004, ShnirmanPRL2005, Galperin2006, paladino_noise_2014,rower_evolution_2023} show typical features of  $1/f^\eta$ noise. This involves strongly coupled low-frequency bath modes, ultraslow relaxation, and often purely incoherent dynamics. Although the microscopic origin of this noise can be difficult to pinpoint, in many cases, environmental quantum two-level fluctuators as encountered in amorphous systems \cite{JaeckleGlaeser1972, NalbachJLTPTS2004} are clearly identified \cite{Lisenfeld15, Lisenfeld19, deGraaf} as its cause. Quantum statistical noise due to a few two-level fluctuators is clearly not Gaussian and cannot be adequately modeled using a spin-boson system with $s>0$~\cite{paladino_noise_2014}. In \ep{many cases}, the theoretical description is restricted to weak coupling or high temperature, employing highly nontrivial approximate methods \cite{Paladino2002,YouPRR2021}. Yet, effects of  $1/f^\eta$ noise sources on the dynamics of single qubits as described, e.g., by the McWhorter model~\cite{McWhorter,weissman} 
are well mimicked by a classical stochastic process, which becomes Gaussian in the limit of many weakly-coupled two-level fluctuators, provided that a low-frequency cutoff is introduced on a phenomenological basis~\cite{falci2005,Ithier2005}.

While the physics is well understood for weakly damped single qubits at high temperatures, much less is known for larger architectures~\cite{kjaergaard_superconducting_2020} where 
quantum features of the low-frequency environment are likely to play an active role. In this respect, the spin-boson model serves as a paradigmatic building block for understanding quantum effects of dissipation for spectral exponents $s<0$ in the roadmap of upscaling to larger systems. This $1/f^\eta$-type quantum noise is particularly interesting at low temperatures. Being notoriously difficult, a numerically exact investigation of the corresponding highly non-Markovian and strongly damped quantum dissipative dynamics had remained elusive. Yet, $1/f^\eta$-type quantum noise has been recognized as a major limiting factor determining dephasing of superconducting qubits due to magnetic flux noise \cite{rower_evolution_2023}. Additionally, the electronic spin of a color center that couples to the vibrational motion of free-standing hexagonal boron nitride membranes has been analyzed in terms of the spin-boson model with $s=-1$ using approximate hybrid numerical tools \cite{Abdi18}.

In this work, we extend the spin-boson model to the regime of quantum  $1/f^\eta$ noise and simulate the dynamics for negative spectral exponents $s<0$ resulting in $1/f^\eta$ noise with $\eta=-s$. We employ the numerically exact time-evolving matrix product operator (TEMPO) technique~\cite{strathearn_efficient_2018,strathearn_modelling_2020}, which is based on the quasiadiabatic propagator path integral (QUAPI) \cite{makri_tensor_1995,makri_tensor_1995-1}. Our investigation yields the dynamic regime diagram of the different regimes of coherent, incoherent and pseudo-coherent  dynamical behavior. Notably, the latter survives for $s<0$, i.e., even in the presence of strong low-frequency bath modes. We demonstrate that the associated dephasing rate still depends linearly on the damping strength for weak damping. Furthermore, we calculate the proportionality factor and find an exponential dependence of the dephasing rate on $s<0$ for zero temperature. As the bath reorganization energy depends sensitively on the infrared cutoff of the fluctuational modes, there is an increased sensitivity of the dephasing rate on the measurement time of an experiment. 

\section{Model}

We consider the symmetric spin-boson model ($\hbar = 1$, $k_B = 1$) with 
the Hamiltonian 
\begin{eqnarray}
\hat{H} &&= \hat{H}_S + \hat{H}_B + \hat{H}_{\text{int}} \nonumber \\
&&=\frac{\Omega}{2} \hat{\sigma}_x + \frac{1}{2} \sum_j \left( \hat{p}_j^2 + \omega_j\hat{x}_j^2 \right)  + \frac{\hat{\sigma}_z}{2} \hat{\xi},
\end{eqnarray}
where $\hat{\xi} = \sum_j c_j \hat{x}_j$, $\Omega$ denotes the tunneling splitting, and $\hat{\sigma}_{x/z}$ are the Pauli matrices. The bath is composed of harmonic oscillators with momenta $\hat{p}_j$, angular frequencies $\omega_j$, and positions $\hat{x}_j$ that are coupled to the spin via coupling constants $c_j$. The bath has the spectral density
\begin{equation} \label{specdens}
    J(\omega ) = \sum_j \frac{c_j^2}{2 \omega_j} \delta (\omega - \omega_j ) = 2 \alpha \frac{\omega^s}{\omega_c^{s-1}} e^{- \frac{\omega}{\omega_c}} \Theta(\omega-\omega_{\text{ir}}),
\end{equation}
with a high-frequency cutoff of $\omega_c = 10 \Omega$ (unless specified otherwise), a low-frequency cutoff $\omega_{\text{ir}}$, and a coupling strength of $\alpha$. We consider spectral exponents $-1 \leq s \leq 1$.
The polarization dynamics $P(t) = \Braket{\sigma_z} (t)$ is calculated assuming factorizing initial preparation of the system with $P(0) = 1$ and the thermal distribution of the initially uncoupled bath at temperature $T$. For the Ohmic and sub-Ohmic baths, other forms of the initial preparation leave the results qualitatively unchanged; quantitative modifications are discussed in Ref.~\cite{Gelin23}. In addition, we calculate the dynamics of the coherence $C(t) = \Braket{\sigma_x} (t)$ as well.

\section{Method}
To determine the time-dependent reduced density matrix, we use the numerically exact TEMPO technique \cite{strathearn_efficient_2018,strathearn_modelling_2020}, which is based on the QUAPI approach \cite{makri_tensor_1995,makri_tensor_1995-1}. The bath oscillators are integrated out analytically, and the remaining path sum representing the highly non-Markovian dynamics of the system is numerically computed by the contraction of a tensor network.
The central quantity is the pair correlation function~\cite{WeissBook}
\begin{eqnarray}
\label{Q_t}
    Q(t) = \frac{1}{\pi} \int_0^{\infty} \mathrm{d}\omega \frac{J(\omega )}{\omega^2} \bigg[ \coth \left( \frac{\omega }{2T} \right) \left( 1- \cos (\omega t)\right) \nonumber \\
    + i \left( \sin (\omega t ) - \omega t\right) \bigg]\, .
\end{eqnarray}
We note that the term $-i\omega t$ has been added. This is allowed because of the freedom of adding an integration constant when going from the original bath autocorrelation function $L(t)$ \cite{WeissBook} to its second integral $Q(t)$. This has convenient consequences, because the parameter range for which the frequency integral in Eq.\ (\ref{Q_t}) is convergent, becomes larger (see below). 
The low-frequency cutoff $\omega_{\text{ir}}$ has been introduced in Eq.\ (\ref{specdens}) to make the integral well-defined for $s \leq 0$ at finite temperature and $s \leq -1$ at zero temperature. 

\begin{figure*}[t!]
\includegraphics[width=170mm]{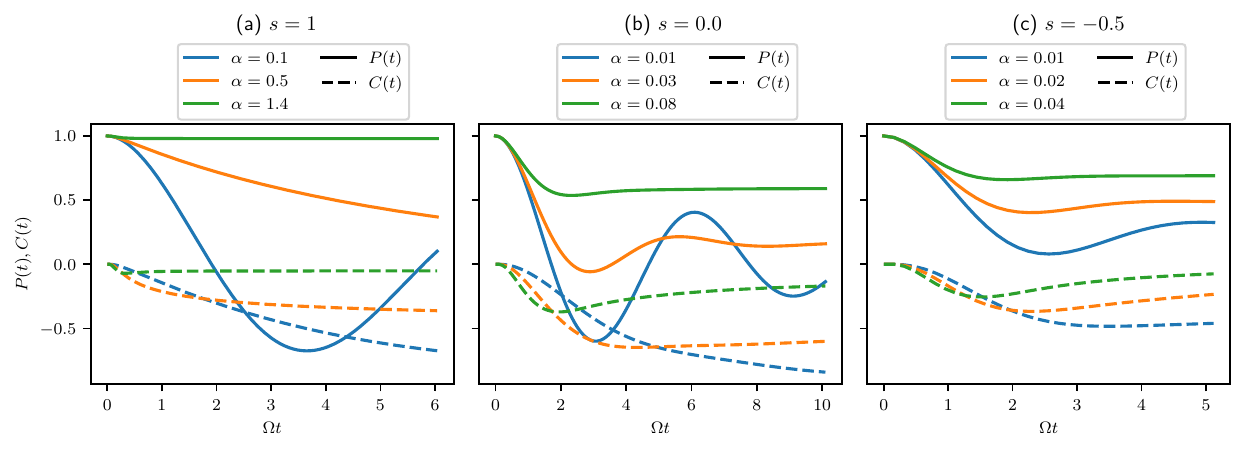}
\caption{Time dependent polarization $P(t)$ (solid lines) and coherence $C(t)$ (dashed lines) for three different choices of the spectral exponent (a) $s=1$, (b) $s=0$, and (c) $s=-0.5$ for $T=0, \omega_c = 10 \Omega$, and $\omega_{\text{ir}}=0$.}%
\label{fig:fig1}
\end{figure*}

\section{Dynamics of the Polarization and Coherence}

In Fig.\ \ref{fig:fig1}, we show the results for the polarization $P(t)$ and the coherence $C(t)$ at zero temperature and no infrared cut-off for three different representative choices of the spectral exponent, i.e., $s=1$, $s=0$, and $s=-0.5$, each for different damping constants $\alpha$ (as indicated) which are chosen to reflect the crossover between the various dynamical regimes. The first case $s=1$ (Fig.\ \ref{fig:fig1} a)) marks the standard Ohmic bath which has been studied in numerous works before. For increasing damping, the polarization dynamics crosses over from coherent damped oscillations (blue line with $\alpha=0.1$) to incoherent decay orange and green lines with $\alpha=0.5$ and $1.4$ respectively. For $s=0$ (Fig.\ \ref{fig:fig1} b)), the time traces are shown for $\alpha \in \lbrace 0.01,0.03,0.08 \rbrace$ and we observe coherent behavior for the two smaller couplings (blue and orange lines) and pseudo-coherent dynamics (green line) at strongest coupling. In the regime of negative spectral exponents, we show the time dependent polarization for $s=-0.5$ and $\alpha \in \lbrace 0.01,0.02,0.04 \rbrace$ in Fig.\ \ref{fig:fig1} c). Here, we find coherent dynamics at $\alpha=0.1$ and pseudo-coherent behavior for the two larger couplings. 

Under weak coupling conditions one expects exponential decay towards thermal equilibrium in the coherence dynamics $C(t)$. This is observed for the Ohmic case and the $s=0$ case at weakest coupling (blue dashed lines). At larger couplings one expects also initial dynamics on a time scale determined by bath parameters followed by decay dynamics. This is clearly observed in the orange and green dashed lines. Notice that the relevant time scale is shortest in the Ohmic case despite using for all simulations the same bath cut-off frequency. 

Remarkably, we observe qualitatively the same dynamical behavior for negative exponents of the spectral density, i.e. $s<0$, as was seen for the deep sub-Ohmic regime, i.e. $0\le s \lesssim 0.45$.

\section{Dynamic Regime Diagram} 

\begin{figure}[t!]
\includegraphics[width=86mm]{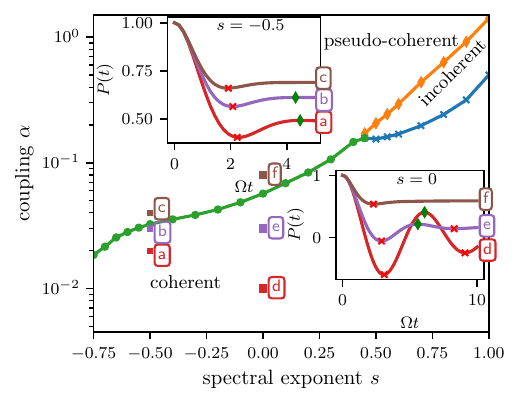}
\caption{ Dynamical regime diagram of the spin-boson model at $T=0, \omega_c = 10 \Omega$, and $\omega_{\text{ir}}=0$. The green line marks the transition from the coherent to pseudo-coherent behavior. The blue line represents the transition from the coherent to incoherent behavior. The orange line represents transition from incoherent to pseudo-coherent behavior. Top left inset: Polarization dynamics for $s=-0.5$ and $\alpha \in \lbrace 0.02,0.03,0.04 \rbrace$ with minima (maxima) marked by a red cross (green diamond) showing the transition from the coherent to the pseudo-coherent dynamical regime. Bottom right inset: Polarization dynamics for $s=0$ and $\alpha \in \lbrace 0.01,0.03,0.08 \rbrace$.}%
\label{fig:1f_phasediagram}
\end{figure}

The polarization dynamics exhibits two distinct regimes: incoherent decay or damped oscillations. The latter regime can be further divided into a coherent and a pseudo-coherent regime. The key difference lies in the oscillation frequency: in the coherent regime, it is determined by both system and bath features, while in the pseudo-coherent regime, it is solely determined by bath parameters. 
 Pseudo-coherent dynamics is observed for system-bath couplings which render the system in the localized (equilibrium) phase. The observed oscillations (due to the non-equilibrium initial condition) are governed solely by the energy scale of the bath ultraviolet cut-off frequency. Thus, it is less the coherence of the central quantum system, but more the coherence of the  bath of harmonic oscillators to which the quantum system is so strongly coupled that its dynamics is enslaved to the short-lived coherent initial bath oscillations.
The complete dynamical regime diagram for the polarization is depicted in Fig.~\ref{fig:1f_phasediagram}.

For spectral exponents $0\leq s \leq 1$, the dynamical regime diagram was determined in Ref.\ 
\cite{otterpohl_hidden_2022} and consists of coherent polarization dynamics at weak coupling, transitioning to incoherent behavior for $s \geq 0.45$ as the coupling strength is increased. Further increasing the coupling strength leads to the pseudo-coherent regime, which is characterized by a single initial oscillation arising from the finite bath reaction time determined by $1/\omega_c$. 
For $s \leq 0.45$, no incoherent dynamical regime appears, and the system directly transitions from the coherent to the pseudo-coherent dynamical regime as the coupling strength is increased. This transition was defined to occur for the coupling strength for which the first local maximum of the coherent oscillations vanishes.

Our results allow us to complete the dynamical regime diagram \cite{footnote} for $s < 0$, as shown in Fig.~\ref{fig:1f_phasediagram}. In particular, we report the dynamical regime diagram in the regime of quantum $1/f^\eta$ noise down to $s=-0.75$, which is the smallest spectral exponent where numerical convergence could be achieved. Interestingly, the transition from coherent to pseudo-coherent dynamics smoothly extends to $-0.75 \leq s \leq 0.45$. Notably, no incoherent dynamical regime emerges in the $1/f^\eta$ regime, contrary to expectations. Instead, the qualitative behavior known from $0 \leq s \leq 0.45$ extends to negative spectral exponents.

The time traces of $P(t)$ for different values of $s$ and $\alpha$ are shown in the insets of Fig.~\ref{fig:1f_phasediagram}. In the top left inset, we depict three representative dynamics for $s=-0.5$: coherent behavior for $\alpha \in \lbrace 0.02, 0.03 \rbrace$, and pseudo-coherent behavior with only one minimum for $\alpha= 0.04$. Similarly, the bottom right inset shows three representatives for $s=0$: coherent dynamics for $\alpha \in \lbrace 0.01,0.03 \rbrace$, and pseudo-coherent behavior for $\alpha= 0.08$.

Remarkably, the dynamical regime diagram reveals an inflection point in the critical coupling strength $\alpha (s)$ at $s \approx -0.5$. We note that we have set $\omega_{\text{ir}}=0$ here due to $T=0$. We conclude that the pseudo-coherent dynamical regime, characterized by a single minimum of $P(t)$ arising on the timescale $1/\omega_c$, persists also for quantum $1/f^\eta$ noise $s<0$.

\section{Dephasing Rate}
\begin{figure}[t!]
\includegraphics[width=86mm]{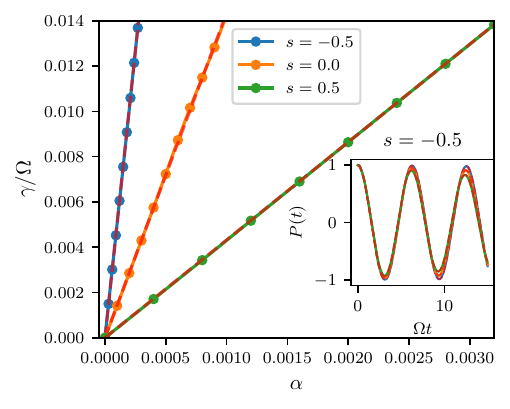}
\caption{Dephasing rate as a function of the coupling strength for $T=0$ for different spectral exponents with linear fits (dashed lines). Inset: Polarization dynamics $P(t)$ for $s=-0.5$ and $\alpha \cdot 10^{4} \in \lbrace 0.3, 1.5, 3\rbrace $ fitted with function given in Eq.~(\ref{weiss_Pt}) (dashed lines). Moreover, $\omega_c = 10 \Omega$, and $\omega_{\text{ir}}=0$.}%
\label{fig:gamma_alpha_fit}
\end{figure}

To gain further insight, we next focus on the regime of small $\alpha$ and use a fitting function~\cite{WeissBook}
\begin{equation}
\label{weiss_Pt}
    P(t) \approx \frac{1}{1 - { \tilde{\alpha}}} \cos ( \omega_0 t ) e^{-\gamma t} - \frac{  { \tilde{\alpha}} }{1 -  { \tilde{\alpha}} } \, .
\end{equation}
In the Ohmic case $s=1$, for $T=0$, and in the scaling limit $\omega_c \rightarrow \infty$, this form follows from the noninteracting blip approximation (NIBA) ~\cite{WeissBook} (up to an incoherent term less relevant at small coupling strengths). In this case, the parameters are given by
\begin{eqnarray} \label{NIBA}
    \omega_0 = \Omega_{\text{eff}} \cos \left[ \frac{\pi \alpha}{2(1-\alpha)} \right] ,\ 
    \gamma = \Omega_{\text{eff}} \sin \left[ \frac{\pi \alpha}{2(1-\alpha)} \right], \nonumber \\
    \frac{\Omega_{\text{eff}}}{\Omega} = \left[ \Gamma(1-2\alpha ) \cos (\pi \alpha) \right]^{1/2(1-\alpha )} \left(\frac{\Omega}{ \omega_c}  \right)^{\alpha / (1-\alpha )}. 
\end{eqnarray}
We find that Eq.~\ref{weiss_Pt} fits our numerical data well across a wide range of spectral exponents and temperatures in the weak coupling regime, using ${ \tilde{\alpha}} , \omega_0$, and $\gamma$ as free parameters. Illustrative fits for the case $s=-0.5$ and zero temperature are shown in the inset of Fig.~\ref{fig:gamma_alpha_fit}. 
From these fits, we extract the dephasing rates $\gamma$ for coupling strengths where the dynamics remains weakly damped, resulting in a linear dependence of $\gamma$ on $\alpha$ (see Fig.~\ref{fig:gamma_alpha_fit}). Notably, this linear dependence persists even in the regime of $1/f^\eta$-type quantum noise, where one might naively expect strongly damped dynamics.

\begin{figure}[t!]
\includegraphics[width=86mm]{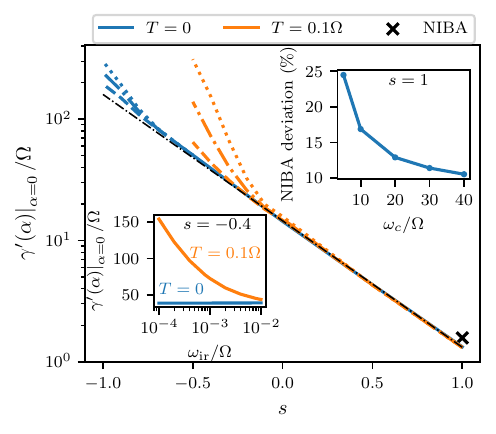}
\caption{Slope $\left. \gamma^\prime (\alpha ) \right|_{\alpha = 0}$ as a function of the spectral exponent for temperatures $T =0$ and $T= 0.1 \Omega$ and low-frequency cutoffs $\omega_{\text{ir}} = 0$ (solid lines), $\omega_{\text{ir}} = 10^{-4}$ (dotted lines), $\omega_{\text{ir}} = 10^{-3}$ (dash-dotted lines), and $\omega_{\text{ir}} = 10^{-2}$ (dashed lines). The thin dash-dotted black line corresponds to Eq.~(\ref{gamma_empirical}). Bottom left inset: $\gamma^\prime (\alpha) |_{\alpha=0}$ as a function of the low-frequency cutoff for $s=-0.4$ and temperatures as indicated.
%
The black cross marks the NIBA result, Eq.\ (\ref{NIBA}). Top right inset: relative deviation of TEMPO from NIBA $1 - \gamma^\prime (\alpha) |_{\alpha=0} / \gamma^\prime (\alpha) |_{\alpha=0}^{\text{NIBA}}$ as a function of $\omega_c$ for $s=1$.}%
\label{fig:k1_vs_s}
\end{figure}

To illustrate the linear dependence $\gamma \propto \alpha$, we compute the proportionality constant $\left. \gamma^\prime (\alpha ) \right|_{\alpha = 0}$ for various spectral exponents and temperatures by performing linear fits on the function $\gamma (\alpha )$. The results are shown in Fig.~\ref{fig:k1_vs_s}.
We find that in the sub-Ohmic regime ($0 < s < 1$), the temperature dependence (at low temperatures, i.e., $T\lesssim\Omega$) is negligible, whereas for $s<0$, the temperature dependence can become significant. Furthermore, despite the divergence of the integral in Eq.\ (\ref{Q_t}) for $s=0$ at finite temperature, the dynamics is largely independent of $\omega_{\text{ir}}$ in the time window considered here (the true asymptotic state is most likely dependent on $\omega_{\text{ir}}$). However, for smaller values of $s$, the dephasing rates can be made arbitrarily large by choosing a correspondingly smaller low-frequency cutoff $\omega_{\text{ir}}$. Thus, the low-frequency cutoff plays a crucial role for dephasing. In any measurement, the low-frequency cutoff is a characteristic essential parameter of the experimental setup as confirmed in the experiment of Ref.~\cite{rower_evolution_2023}. 

We empirically find the dephasing rate
\begin{equation}
\label{gamma_empirical}
 \gamma \approx 14.5 \, \alpha \, e^{-2.4 s} \, { \Omega }
\end{equation}
from the data in Fig.~\ref{fig:k1_vs_s} for $\alpha \ll 1$ and $\omega_{\text{ir}} = 0$. It is valid for $s \gtrsim 0$ at finite temperatures and for $s \gtrsim -0.5$ at zero temperature. The deviation of the numerical results from the NIBA value of $\gamma^\prime (\alpha) |_{\alpha=0}^{\text{NIBA}} = \Omega \pi/2$ for $s=1$ arises from the finite high-frequency cutoff (see top right inset of Fig.~\ref{fig:k1_vs_s}). Thus, while Eq.~(\ref{gamma_empirical}) has been obtained from data with $\omega_c = 10\Omega$, it also yields an approximation of the dephasing rate for a wide range of $\omega_c$ due to the weak dependence of the dynamics on $\omega_c$ at weak coupling.

\section{Reorganization energy}
\begin{figure}[t!]
\includegraphics[width=86mm]{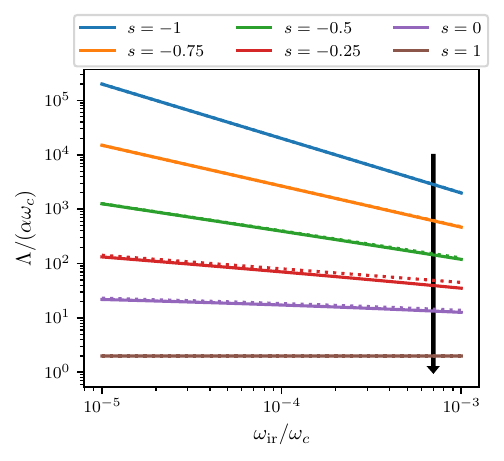}
\caption{Reorganization energy $\Lambda$ as a function of $\omega_{\text{ir}} / \omega_c$ for $s \in \lbrace -1,-0.75,-0.5,-0.25,0,1 \rbrace$. The solid lines show the exact result for $\Lambda$ of 
Eq.~(\ref{reorg}), while the dotted lines show the approximation for $\omega_{\text{ir}} \ll \omega_c$ given in the text. The arrow intersects the lines in ascending magnitude of $s$.}%
\label{fig:reorganization_energy}
\end{figure}

To quantify the energy contained in the bath fluctuations, we calculate the reorganization energy  \cite{WeissBook}, given by
\begin{equation}
\label{reorg}
 \Lambda = \int_{\omega_{\text{ir}}}^\infty d\omega \frac{J(\omega)}{\omega} 
 = 2 \alpha \omega_c \Gamma(s, \omega_{\text{ir}} / \omega_c)\, .
\end{equation}
Here, $\Gamma(s,x)$ is the upper incomplete Gamma function. In the limit of $\omega_{\text{ir}} \ll \omega_c$, we find that 
$\Lambda  = -2\alpha\omega_c(\omega_{\text{ir}}/\omega_c)^s/s$ for $s < 0$, which diverges for $\omega_{\text{ir}} \rightarrow 0$. For $s = 0$, we have $\Lambda  =-2\alpha\omega_c\ln (\omega_{\text{ir}}/\omega_c)$, and for 
$s > 0$, we find $\Lambda  =2\alpha\omega_c \Gamma(s)$, where $\Gamma (s)$ denotes the gamma function. The corresponding behavior is shown in Fig.~\ref{fig:reorganization_energy}. 
%
Surprisingly, as shown in Fig.~\ref{fig:k1_vs_s}, we find that the polarization dynamics is well-defined in the limit $\omega_{\text{ir}} \rightarrow 0$ for at least some $s<0$ at zero temperature.

An experiment inherently implies a finite measurement time window $t_{\text{meas}}$ which motivates a minimal infrared cutoff $\omega_{\text{ir}}=2\pi / t_{\text{meas}}$. Hence, the longer the data acquisition occurs, the stronger the impact of the quantum $1/f^\eta$ noise is, which is consistent with Ref.\ \cite{rower_evolution_2023}. 

\section{Conclusion}
Using the numerically exact TEMPO approach, we have investigated the dynamics of the spin-boson model in the $1/f^\eta$ quantum noise regime, where strong low-frequency bath modes exist at zero temperature. Our analysis classifies the polarization dynamics into three regimes: coherent, incoherent, and the pseudo-coherent. In the latter, the dynamics is dominated by the coherent dynamics of the bath, rendering it independent of the system energy scale. Consequently, the system dynamics is effectively overdamped. From the polarization dynamics, we extract the dephasing rate at weak system-bath coupling for spectral exponents deep into the $1/f^\eta$ regime for $s<0$. We provide an empirical formula for the dephasing rate at zero temperature across the full parameter space, which remains valid at finite temperature for a significant portion of the studied parameter range. 
\ep{Moreover}, we reveal the sensitivity of the dissipative dynamics on the infrared cutoff \ep{with no need of phenomenological assumptions}, which explains the strong dependence of the dephasing on the observation time in flux-noise-limited dephasing in superconducting qubits. \ep{We finally point out that path-integral techniques used in this work, differently than approximate master equations, provide a general tool to study open quantum systems also in the ``Brownian''  regime of 
low-temperatures or decoherence rates comparable to the natural splittings~\cite{Ankerhold_2003}. Therefore,} we envision that this work potentially helps to provide designing guidelines for the realization of \ep{multiqubit architectures based on} superconductors, \ep{where the presence of several time scales \ep{and crowded spectral densities \cite{Plourde_2017}} hinders making clear phenomenological assumptions}.

\begin{acknowledgments}
\ep{GF acknowledges support from the ICSC – Centro Nazionale di Ricerca in High-Performance Computing, Big Data and Quantum Computing; EP is supported by the
PNRR MUR project PE0000023-NQSTI;   GF and EP acknowledge support from the University of Catania, Piano Incentivi Ricerca di Ateneo 2024-26, project
QTCM. EP acknowledges the COST Action SUPERQUMAP (CA 21144).}
\end{acknowledgments}


\begin{thebibliography}{50}

\bibitem{WeissBook} U. Weiss, Quantum Dissipative Systems, 5th ed. (World Scientific, Singapore, 2021).

\bibitem{Leggett1987} A.~J. Leggett, S. Chakravarty, A.~T. Dorsey, M.~P.~A. Fisher, A. Garg, and W. Zwerger, Rev. Mod. Phys. {\bf 59}, 1 (1987).

\bibitem{otterpohl_hidden_2022} F. Otterpohl, P. Nalbach, and M. Thorwart, Hidden phase of the spin-boson model, Phys. Rev. Lett. {\bf 129}, 120406 (2022).


\bibitem{Chin06} A. Chin and M. Turlakov, Coherent-incoherent transition in the sub-Ohmic spin-boson model, Phys. Rev. B {\bf 73}, 075311 (2006).

\bibitem{Anders07} F. B. Anders, R. Bulla, and M. Vojta, Equilibrium and
nonequilibrium dynamics of the sub-Ohmic spin-boson
model, Phys. Rev. Lett. {\bf 98}, 210402 (2007).

\bibitem{Winter09} A. Winter, H. Rieger, M. Vojta, and R. Bulla, Quantum phase transition in the sub-Ohmic spin-boson model: Quantum Monte Carlo study with a continuous imaginary time cluster algorithm, Phys. Rev. Lett. {\bf 102},
030601 (2009).

\bibitem{Alvermann09} A. Alvermann and H. Fehske, Sparse polynomial space
approach to dissipative quantum systems: Application to the sub-Ohmic spin-boson model, Phys. Rev. Lett. {\bf 102}, 150601 (2009).

\bibitem{Nalbach10} P. Nalbach and M. Thorwart, Ultraslow quantum dynamics in a sub-Ohmic heat bath, Phys. Rev. B {\bf 81}, 054308 (2010).

\bibitem{Wang10}H. Wang and M. Thoss, From coherent motion to localization: II. Dynamics of the spin-boson model with sub-Ohmic spectral density at zero temperature, Chem. Phys. {\bf 370}, 78 (2010).

\bibitem{Chin11}A. W. Chin, J. Prior, S. F. Huelga, and M. B. Plenio, Generalized polaron ansatz for the ground state of the sub-Ohmic spin-boson model: An analytic theory of the localization transition, Phys. Rev. Lett. {\bf 107}, 160601 (2011).

\bibitem{Kast13} D. Kast and J. Ankerhold, Persistence of coherent quantum dynamics at strong dissipation, Phys. Rev. Lett. {\bf 110}, 010402 (2013).

\bibitem{Nalbach13} P. Nalbach and M. Thorwart, Crossover from coherent to in coherent quantum dynamics due to purely dephasing Sub-Ohmic fluctuations, Phys. Rev. B {\bf 87}, 014116 (2013).

\bibitem{Schroeder16} F. A. Y. N. Schröder and A. W. Chin, Simulating open quantum dynamics with time-dependent variational matrix product states: Towards microscopic correlation of environment dynamics and reduced system evolution, Phys. Rev. B {\bf 93}, 075105 (2016). 

\bibitem{Duan17} C. Duan, Z. Tang, J. Cao, and J. Wu, Zero-temperature localization in a sub-Ohmic spin-boson model investigated by an extended hierarchy equation of motion, Phys. Rev. B {\bf 95}, 214308 (2017).

\bibitem{Gelin23} L. Chen, Y. Yan, M.F. Gelin, and Z. Lü, Dynamics of the spin-boson model: The effect of bath initial conditions, J. Chem. Phys. {\bf 158}, 104109 (2023).


\bibitem{Falci2024} G. Falci, Pertti J. Hakonen, and E. Paladino, {\em $1/f$ noise in quantum nanoscience}, in {\em Encyclopedia of Condensed Matter Physics (2nd ed.)}, edited by T. Chakraborty (Academic Press, Cambridge, MA, 2024),  pp. 1003-1017. 


\bibitem{Paladino2002} E. Paladino, L. Faoro, G. Falci, and R. Fazio, Decoherence and $1/f$ noise in Josephson qubits, Phys. Rev. Lett. {\bf 88}, 228304 (2002).

\bibitem{AstafievPRL2004} O. Astafiev, Yu. A. Pashkin, Y. Nakamura, T. Yamamoto, and J. S. Tsai, Quantum noise in the Josephson charge qubit, Phys. Rev. Lett. {\bf 93}, 267007 (2004). 

\bibitem{ShnirmanPRL2005} A. Shnirman, G. Schön, I. Martin, and Yuriy Makhlin, Low- and high-frequency noise from coherent two-level systems, Phys. Rev. Lett. {\bf 94}, 127002, (2005).

\bibitem{Galperin2006} Y. M. Galperin, B. L. Altshuler, J. Bergli, D. V. Shantsev, Non-Gaussian low-frequency noise as a source of qubit decoherence, Phys. Rev. Lett. {\bf 96}, 097009 (2006).

\bibitem{paladino_noise_2014} E. Paladino, Y. M. Galperin, G. Falci, and B. L. Altshuler, $1/f$ Noise: Implications for solid-state quantum information, Rev. Mod. Phys. {\bf 86}, 361 (2014).

\bibitem{rower_evolution_2023} D. A. Rower et al., Evolution of $1/f$ flux noise in superconducting qubits with weak magnetic fields, Phys. Rev. Lett. {\bf 130}, 220602 (2023).


\bibitem{JaeckleGlaeser1972} J. Jäckle, On the ultrasonic attenuation in glasses at low temperatures, Z. Physik {\bf 257}, 212 (1972).

\bibitem{NalbachJLTPTS2004} P. Nalbach, D. Osheroff, and S. Ludwig,  Non-equilibrium dynamics of interacting tunneling states in glasses, J. Low Temp. Phys. {\bf 137}, 395 (2004). 

\bibitem{Lisenfeld15} J. Lisenfeld, G. J. Grabovskij, C. Müller, J. H. Cole, G. Weiss, and A. V. Ustinov, Observation of directly interacting coherent two-level systems in a solid, Nat. Commun. {\bf 6}, 6182 (2015).

\bibitem{Lisenfeld19} C. Müller, J. H. Cole, and J. Lisenfeld, Towards understanding two-level-systems in amorphous solids: insights from quantum circuits, Rep. Prog. Phys. {\bf 82}, 124501 (2019).

\bibitem{deGraaf} S. E. de Graaf, L. Faoro, L. B. Ioffe, S. Mahashabde, J. J. Burnett, T. Lindström, S. E. Kubatkin, A. V. Danilov, and A. Ya. Tzalenchuk, Two-level systems in superconducting quantum devices due to trapped quasiparticles, Science Advances {\bf 6}, eabc5055 (2020).

\bibitem{YouPRR2021} X. You, A. A. Clerk, and J. Koch, Positive- and negative-frequency noise from an ensemble of two-level fluctuators, Phys. Rev. Research {\bf 3}, 013045 (2021).

\bibitem{McWhorter} \ep{A.L. McWhorter, in Proceedings of the Conference on Physics of Semiconductor Surface Physics, R.H. Kingston (ed.), University of Pennsylvania Press, Pennsylvania (1957), p. 207. }%

\bibitem{weissman} \ep{M.B. Weissman, 1/f noise and other slow, nonexponential kinetics in condensed matter, Rev. Mod. Phys. {\bf 60}, 537 (1988).}

\bibitem{falci2005}\ep{G. Falci, A. D’Arrigo, A. Mastellone, and E. Paladino,
Initial Decoherence in Solid State Qubits, Phys. Rev. Lett. {\bf 94}, 167002, (2005).}

\bibitem{Ithier2005}\ep{G. Ithier, E. Collin, P. Joyez, P. J. Meeson, D. Vion, D. Esteve, F. Chiarello, A. Shnirman, Y. Makhlin, J. Schriefl, and G. Schön, 
Decoherence in a superconducting quantum bit circuit, 
Phys. Rev. B {\bf 72}, 134519 (2005).}

\bibitem{kjaergaard_superconducting_2020} 
\ep{M. Kjaergaard, M. E. Schwartz, J. Braumüller, P. Krantz, J. I-Jan Wang, S. Gustavsson, and W. D. Oliver, Superconducting Qubits: Current State of Play, Annu. Rev. Condens. Matter Phys. {\bf 11}, 369 (2020).}

\bibitem{Abdi18} M. Abdi and M. B. Plenio, Analog quantum simulation of extremely sub-Ohmic spin-boson models, Phys. Rev. A {\bf 98}, 040303(R) (2018).







\bibitem{strathearn_efficient_2018} A. Strathearn, P. Kirton, D. Kilda, J. Keeling, and B. W. Lovett, Efficient non-Markovian quantum dynamics using time-evolving matrix product operators, Nat. Commun. {\bf 9}, 3322 (2018).

\bibitem{strathearn_modelling_2020} A. Strathearn, Modelling non-Markovian quantum systems using tensor networks (Springer International Publishing, Cham, 2020).

\bibitem{makri_tensor_1995} N. Makri and D. E. Makarov, Tensor propagator for iterative quantum time evolution of reduced density matrices. I. Theory, J. Chem. Phys. {\bf 102}, 4600 (1995).

\bibitem{makri_tensor_1995-1} N. Makri and D. E. Makarov, Tensor Ppopagator for iterative quantum time evolution of reduced density matrices. II. Numerical methodology, J. Chem. Phys. {\bf 102}, 4611 (1995).

\bibitem{footnote} We note that the dynamical regime diagram is often denoted as phase diagram in the literature. 


\bibitem{Ankerhold_2003} \ep{J. Ankerhold, Dynamics of dissipative quantum systems -- from path integrals to master equations, Lecture Notes in Physics vol.\ 622, F. Benatti, R. Floreanini (eds.) (Springer, New York, 2003), p.\ 165.}



\bibitem{Plourde_2017} \ep{M.D. Hutchings, J.B. Hertzberg, Y. Liu1, N.T. Bronn, G.A. Keefe, M. Brink, J. M. Chow, and B.L.T. Plourde, 
Tunable Superconducting Qubits with Flux-Independent Coherence, Phys. Rev. Applied {\bf 8}, 044003 (2017).}





\end{thebibliography}
\end{document}